\documentclass[12pt]{amsart}
\usepackage{a4wide,palatino,epsfig,latexsym,
amsmath,amsthm,graphicx,hyperref,enumerate,morefloats,color, natbib, soul}
\usepackage{mathtools}

\title{Walsh coefficients and circuits for several alleles}

  \author{Kristina Crona and Devin Greene}

\theoremstyle{plain}

\theoremstyle{definition}

\begin{document}

\maketitle

\begin{abstract}
Walsh coefficients have been applied extensively 
to biallelic systems for quantifying 
pairwise and higher order epistasis,
in particular for demonstrating
the empirical importance of
higher order interactions.
Circuits, or minimal dependence 
relations, 
and related approaches
that use triangulations
of polytopes
have also been applied
to biallelic systems.
Here we 
provide biological interpretations of
Walsh coefficients for several alleles,
and discuss circuits in the same general
setting.
\end{abstract}

\section{Introduction}
Walsh coefficients are useful for analyzing epistasis,
or deviations from additive fitness \citep{wlw}.
Extension of Walsh coefficients
to an arbitrary number of alleles
have been discussed in recent work \citep{gs, mps, flm}.
Here, we demonstrate that
Walsh coefficients for several
alleles are easy to interpret
as deviations from additivity,
similar to the biallelic case.
We also discuss circuits  \citep{bps} for 
several alleles.
The paper is self-contained,
no prior knowledge of Walsh coefficients
or circuits is expected.
We start with  interpretations of Walsh coefficients
for biallelic systems. 
In subsequent sections
we analyze several alleles.

\subsection{Biallelic two-locus systems}
Let $w_g$ denote the fitness of a genotype $g$.
By definition, additive fitness
implies that the effect of a double mutation
equals  the sum of the effects of the single mutations it combines.
In other words,
\begin{equation}
w_{11}-w_{00}=(w_{10}-w_{00})+(w_{01}-w_{00})  \label{c0}
\end{equation}
where $00$ denotes the wild-type,
$10$ and $01$ the single mutants,
and $11$ the double mutant.
Equivalently, fitness is additive if
\[
w_{11}=w_{10}+w_{01}-w_{00}.
\]
The form
\[
w_{11}-w_{10}-w_{01}+w_{00}
\]
is referred to as a Walsh coefficient, specifically
the Walsh coefficient of order two.
The form measures epistasis, or
deviations from additivity.
There are in total four Walsh coefficients for the
biallelic two-locus case.
The Walsh coefficient of order zero 
\[
w_{11}+w_{10}+w_{01}+w_{00},
\]
 can be considered a  measure average fitness
(scale factors are ignored here).

The Walsh coefficients of order one 
measure the effect of changing the first locus from 0 to 1:
\[
(w_{10}-w_{00})+(w_{11}-w_{01}),
\]
and of changing the second locus from 0 to 1:
\[
(w_{01}-w_{00})+(w_{11}+w_{10}).
\]

The four Walsh coefficients
\begin{align*}
&  w_{00} + w_{01} + w_{10} + w_{11}  \\
&  w_{00} - w_{01} + w_{10} -  w_{11} \\
&  w_{00} + w_{01} - w_{10} - w_{11}  \\
&  w_{00} - w_{01} - w_{10} +w_{11} 
\end{align*}
can be expressed in a compact way using matrix notation
\[
\begin{bmatrix*}[r]
 1  &1  & 1   &  1      \\
 1  &-1  & 1   &  -1     \\
 1  & 1  & -1   & - 1    \\ 
 1  & -1  &  -1   &  1        
\end{bmatrix*}
\quad
\begin{bmatrix}
 w_{00}   \\
 w_{01}    \\
 w_{10}   \\ 
w_{11}     
\end{bmatrix}.
\]
The matrix is known as the Hadamard matrix. 
In some cases scale factors are included
in the definition. However, we ignore scale factors throughout 
the paper. Walsh coefficients 
provide a compact summary of
average fitness, first order effects (the effect of replacing
the allele at a particular locus) and interactions
for any biallelic $L$-locus
system.

\subsection{Biallelic three-locus systems}
Throughout the paper we will refer to the first, second,
and third locus, and so forth, moving from left to right.
The Hamming distance for a pair of genotypes 
is defined as the number of loci where the
genotypes differ.

The Walsh coefficients for three-locus biallelic systems 
are similar to the two-locus case except that 
the second order interactions should be understood as average
effects over all backgrounds, and 
that there are also  third-order interactions. 
In more detail, as in the two-locus case one can compare
the effect of a pair of  mutations with the effects of each mutation
individually.
Additive fitness implies that
\begin{equation}\label{c31}
w_{110}-w_{000}=(w_{100}-w_{000})+(w_{010}-w_{000})
\end{equation}
Notice that equation is identical to \ref{c0} except for that there is a third
coordinated fixed at zero.
Similarly, if the third locus is fixed at 1 additivity implies that
\begin{equation}\label{c32}
w_{111}-w_{001}=(w_{101}-w_{001})+(w_{011}-w_{001})
\end{equation}
From \ref{c31} and \ref{c32}, additive fitness implies that
\[
w_{110} +w_{111}= w_{000}-w_{001}+w_{010}+w_{011}+w_{100}+w_{101}
\]
Consequently, the difference 
\[
w_{110} +w_{111}-(w_{000}-w_{001}+w_{010}+w_{011}+w_{100}+w_{101})
\]
measures epistasis for the first two loci averaged over all backgrounds ($0$ and $1$),
ignoring scale factors.
In other words, the form \begin{equation}\label{w1}
w_{000}+w_{001}-w_{010}-w_{011}-w_{100} -w_{101}+w_{110} + w_{111} 
\end{equation}
is a measure of two-way epistasis for the first two loci.

By keeping the second locus fixed one derives equations similar to \ref{c31} and \ref{c32}
\begin{align*} 
& w_{101}-w_{000}=( w_{100}-w_{000})+(w_{001}-w_{000} )  \\
& w_{111}- w_{010}=(w_{110}-w_{010})+ (w_{011} - w_{010}).
\end{align*}
From the equations one derives a measure of pairwise epistasis for the first and third locus
averaged over all backgrounds
\begin{equation}\label{w2}
w_{000}-w_{001}+w_{010}-w_{011}-w_{100} +w_{101}-w_{110} + w_{111},
\end{equation}
and similarly for the second and third locus 
\begin{equation}\label{w3}
w_{000}-w_{001}-w_{010}+w_{011}+w_{100} -w_{101}-w_{110} + w_{111}.
\end{equation}
The forms \ref{w1}, \ref{w2}, and \ref{w3} are the second order Walsh coefficients.

It remains to consider third-order interactions. 
If one wants to base a prediction of $w_{111}$
on the wild-type, single mutants and 
pairwise effects, then one needs three correction terms
for the pairwise effects.
By fixing the third coordinate at $0$ one derives
a correction term by noting that additive fitness
implies
\[
w_{110}-w_{000}=(w_{100}-w_{000})+(w_{010}-w_{000}),
\]
or
\[
w_{110}=w_{100}+w_{010}-w_{000}.
\]
It follows that the correction based on pairwise epistasis for the first two loci is
\[
\delta_{12}= w_{110}-w_{100}-w_{010}+w_{000}.
\]
Similarly, the correction term for the first and third loci is
\[
\delta_{13} =w_{101}-w_{100}-w_{001}+w_{000},
\]
and for the second and third loci
\[
\delta_{23} =w_{011}-w_{010}-w_{001}+w_{000}.
\]
A prediction of $w_{111}$ based on $w_{000}$, the effect of the single mutations,
and the correction terms based on pairwise epistasis can be expressed as
\[
w_{111}-w_{000}=(w_{100}-w_{000})+(w_{010}-w_{000})+(w_{001}-w_{000})+ \delta_{12} +\delta_{13}+\delta_{23}
\]
or
\[
w_{111}=w_{110}+w_{101}+w_{011}-w_{100}-w_{010}-w_{001}+w_{000}
\]
Consequently, the form
\begin{equation}\label{w33}
w_{000}-w_{001}-w_{010}+w_{011}-w_{100} +w_{101}+w_{110} - w_{111}
\end{equation}
is a measure of three-way epistasis.

There eight Walsh coefficients for three loci are
\begin{align*}
&  w_{000} + w_{001} + w_{010} + w_{011} + w_{100} + w_{101}+w_{110} + w_{111}  \\
&  w_{000} - w_{001} + w_{010} -  w_{011} + w_{100}  - w_{101}+ w_{110} - w_{111}  \\
&  w_{000} + w_{001} - w_{010} - w_{011} + w_{100} + w_{101} - w_{110} - w_{111}  \\
&  w_{000} -w_{001} - w_{010} +w_{011} +w_{100} - w_{101}- w_{110} + w_{111}  \\
&  w_{000} + w_{001} + w_{010} + w_{011} - w_{100} -w_{101}-w_{110} - w_{111}  \\
&  w_{000} - w_{001} + w_{010} - w_{011} - w_{100} +w_{101}-w_{110} + w_{111}  \\
&  w_{000} + w_{001} -w_{010} -w_{011} - w_{100} -w_{101}+w_{110} + w_{111}  \\
& w_{000} - w_{001} - w_{010} +w_{011} -w_{100} +w_{101}+w_{110} -w_{111} ,
\end{align*}
or in matrix notation
\[
\begin{bmatrix*}[r]
 1  &1  & 1   &  1  &  1  & 1  & 1 &1       \\
 1  &-1  & 1   &  -1  &  1  & -1  & 1 &-1    \\
 1  & 1  & -1   & - 1  &  1  & 1  & - 1 & -1    \\ 
 1  & -1  &  -1   &  1  & 1  & -1  & -1 &1    \\ 
 1  & 1  & 1   &  1  &  -1  & -1  & -1 &-1    \\ 
 1  & -1  & 1   & - 1  &  -1  & 1  & -1 &1    \\ 
 1  & 1  & -1   &  -1  &  -1  & -1  & 1 &1    \\ 
 1  & -1  & -1   &  1  &  -1  & 1  & 1 &-1    
\end{bmatrix*}
\quad
\begin{bmatrix}
 w_{000}   \\
 w_{001}    \\
 w_{010}   \\ 
w_{011}     \\ 
 w_{100}     \\ 
 w_{101}   \\ 
 w_{011}   \\ 
 w_{111}      
\end{bmatrix}
\]
As in the two-locus case,
the Walsh coefficient of order zero (the top tow) is the sum of all genotypes.
The Walsh coefficients of order one (Rows 2, 3 and 5) measure the
 average effect of changing one of the loci (from 0 to 1).
In particular, the  effect of changing the first locus from 0 to 1  is
\[
(w_{100}-w_{000})+ (w_{110}-w_{010}) + (w_{101}-w_{001} ) + (w_{111}-w_{011})=
\]
\begin{equation}\label{m1}
-w_{000} -w_{001} -w_{010}-w_{011}  +  w_{100}  +w_{101}+w_{110}+w_{111},
\end{equation}
and  \ref{m1} is the negative of Row 5.
Similarly,  the second coefficient measures the same effect for the third locus, and Row 3 for the second locus.
Row 4 (see \ref{w3}), Row 6 (see \ref{w2})  and  Row 7 (see \ref{w1})  measure pairwise epistasis.
Row 8 is  the three-way interactions (see \ref{w33}). 

For any $L$ Walsh coefficients summarize average fitness, first order effects (the effect of changing an allele from 0 to 1),
and interactions. In particular, the second and higher order Walsh coefficients are zero if fitness is additive.
The Walsh coefficient $W_{i_1 i_2 \dots i_L}$ for $i_1, \dots i_L \in \{ 0,1 \}$  
is defined as 
\[
W_{i_1 i_2 \dots i_L}  =  %\sum_{0 \leq j_1 \leq 1}  \sum_{0 \leq j_2 \leq 1}  \dots  \sum_{0 \leq j_L \leq 1}  
\sum_{j_1 \dots j_L}  (-1)^{ i_1 j_1 + \dots + i_L  j_L} \, w_{j_1 \dots j_L},
\]
where the sum is taken over all bit-strings  $j_1 \dots j_L$ of length $L$.
For complete proofs see e.g. \citet{gp}.

\subsection{Circuits and Walsh coefficients}
Suppose that fitness is additive with the two exceptions
that $w_{110}$ is slightly higher than additive fitness would imply,
and  $w_{111}$ slightly lower, say
\[
w_{110}-w_{100}-w_{010}+w_{000}= \epsilon
\]
and
\[
w_{111}-w_{101}-w_{011}+w_{001}= -\epsilon
\]
Then the two-way epistasis for the first two loci measured by the Walsh coefficient \ref{w1}
is zero. The reason is that  the positive and negative effects cancel out.
The epistasis will however show up as a third-order effect:
\[
w_{000} - w_{001} - w_{010} +w_{011} -w_{100} +w_{101}+ w_{110} -w_{111}=-2 \epsilon .
\]
For a analyzing gene interactions  on the most detailed level, 
as opposed to considering average effects,
it can be beneficial to consider circuits, i.e., minimal dependence
relations such as
\begin{align*}
w_{110}-w_{100}-w_{010}+w_{000}  \\
w_{111}-w_{101}-w_{011}+w_{001}
\end{align*}

By definition, circuits are linear forms that 
are zero whenever fitness is additive,
and whose support
(i.e., the $w_g$ with non-zero coefficients) 
is non-empty but minimal with respect to inclusion.
There are in total 20 circuits for a  biallelic three-locus system \citep{bps}.
Approaches that depend on circuits have been used in 
different areas of biology. For some some recent 
applications see \cite{ejl, cks, by, c1, eble}.
Circuits for several alleles will be discussed in the next section.

\section{Two loci and three alleles}
The  genotypes for two loci and three alleles are denoted
\[
00, 01, 02, 10, 11, 12, 20, 21, 22.
\]
The analysis of gene interactions is similar to
the biallelic case. One can compare $w_{11}$ 
with $w_{00}$ but it is equally natural to
compare with other genotypes of Hamming distance two,
i.e., $20, 02$ and  $22$.
If one compares with $00$, additive fitness implies
\[
w_{11}-w_{00}= (w_{10}-w_{00}) +(w_{01}-w_{00}),
\]
or
\begin{equation}\label{w23a}
w_{11}= w_{10}+w_{01}-w_{00} 
\end{equation}
If one compares with 22, additive fitness implies
\[
w_{11}-w_{22}=(w_{12}-w_{22}) +(w_{21}-w_{22}),
\]
or 
\begin{equation}\label{w23b}
w_{11}=w_{12}+w_{21}-w_{22}
\end{equation}

By a similar argument for comparisons with 20 and 02
one gets
\begin{equation}\label{w23c}
w_{11}=w_{21}+w_{10}-w_{20}
\end{equation}
and
\begin{equation}\label{w23d}
 w_{11}=w_{12}+w_{01}-w_{02}.
\end{equation}

From \ref{w23a}, \ref{w23b}, \ref{w23c} and \ref{w23d},
additive fitness implies that
\[
w_{11} = \frac{1}{4} \left( 2w_{10}+2w_{01}+2w_{12}+2w_{21}-w_{00}-w_{22}-w_{02}-w_{20} \right)
\]

\[
= \frac{1}{2}  \left( w_{10}+w_{01}+w_{12}+w_{21}\right)  - \frac{1}{4} \left(w_{00}+w_{22}+w_{02}+w_{20} \right).
\]

In other words, epistasis can be measured by the form
\begin{equation}\label{wc231}
w_{11}+1/4 (w_{00}+w_{22}+w_{20}+w_{02}) - 1/2  \left( w_{10}+w_{01}+w_{12}+w_{21} \right).
\end{equation}

A similar analysis of  $w_{12}$, $w_{21}$ and $w_{22}$  give another four measures of epistasis

\begin{equation}\label{wc232}
w_{12}+1/4 (w_{00}+w_{01}+w_{20}+w_{21}) - 1/2  \left( w_{10}+w_{11}+w_{02}+w_{22} \right)
\end{equation}

\begin{equation}\label{wc233}
w_{21}+1/4 (w_{00}+w_{02}+w_{10}+w_{12}) - 1/2  \left(w_{01}+w_{11}+w_{20}+w_{22} \right) 
\end{equation}

\begin{equation}\label{wc234}
w_{22}+1/4 (w_{00}+w_{01}+w_{10}+w_{11}) - 1/2  \left(w_{20}+w_{21}+w_{02}+w_{12} \right).
\end{equation}

The argument applies to $w_{00}$, $w_{10}$, $w_{01}$, $w_{20}$, $w_{02}$ as well. However,
the latter forms can also be expressed as  linear combinations of \ref{wc231},  \ref{wc232}, \ref{wc233} and \ref{wc233}.
Specifically,

\[
w_{00} + \frac{1}{4}(w_{11} + w_{12} +w_{21}+w_{22})-\frac{1}{2}(w_{01} +w_{02}+w_{10}+w_{20}) 
=(13)+(14)+(15)+(16)
%=  \ref{wc231}+ \ref{wc232}+\ref{wc233}+\ref {wc234} 
\]

\[
w_{10} +  \frac{1}{4}(w_{01}+w_{02}+w_{21}+w_{22}) - \frac{1}{2} (w_{11}+w_{12}+w_{00}+w_{02}) 
%=-\ref{wc231}-\ref{wc232}   
=-(13)-(14)
\]

\[
w_{20} +  \frac{1}{4} (w_{01}+w_{02}+w_{11}+w_{22})  - \frac{1}{2} (w_{21}+w_{22}+w_{00}+w_{10})  
=-(15)-(16)
% =-\ref{wc233}-\ref{wc234} 
\]

\[
w_{01}+   \frac{1}{4} (w_{10}+w_{12}+w_{20}+w_{22}) - \frac{1}{2}(w_{00}+w_{02}+w_{11}+w_{12})  
=-(13)-(15)
%= -\ref{wc231}-\ref{wc233} 
\]

\[
w_{02} + \frac{1}{4}( w_{10}+w_{11}+w_{20}+w_{21}) -\frac{1}{2}(w_{01}+w_{02}+ w_{12}+w_{22})   
=-(14)-(16)
%=-\ref{wc232}-\ref{wc234} 
\]

Following \citep{gs}, we consider \ref{wc231},  \ref{wc232},  \ref{wc233} and  \ref{wc234} 
Walsh coefficients. Similar to the biallelic case, all second order interactions 
can be described as linear combinations of the Walsh coefficients.
However, in contrast to the biallelic case there is no obvious preferred
way to define Walsh coefficients (in that sense 
the definition of (13)-(16) as Walsh coefficient is arbitrary, see also the discussion section).

The analysis of lower order  Walsh coefficients is also similar to
the biallelic case. The coefficient of order zero is
\[
w_{00}+w_{01}+w_{02}+w_{10}+w_{20}+w_{11}+w_{12}+w_{21}+w_{22}.
\]
The effect of changing the first locus to $1$ from an alternative allele (0 or 2)
can be measured by the form
\[
(w_{10}-w_{20})+ (w_{10}-w_{00}) + (w_{11}-w_{21} ) + (w_{11}-w_{01})+  (w_{12}-w_{02})+ (w_{12}-w_{22})=
\]
\[
2w_{10}+ 2w_{11}+ 2w_{12} -(w_{00}+w_{01}+w_{02} +w_{20}+w_{21}+ w_{22})
\]
Similarly, the effect of changing the first locus to $2$ from an alternative allele can be measured by the form
\[
2w_{20}+ 2w_{21}+ 2w_{22} -(w_{00}+w_{01}+w_{02} +w_{10}+w_{11}+ w_{12}).
\]
Analogously, for the second locus one gets the two forms
\[
2w_{01}+ 2w_{11}+ 2w_{21} -(w_{00}+w_{10}+w_{20} +w_{02}+w_{12}+ w_{22})
\]
\[
2w_{02}+ 2w_{12}+ 2w_{22} -(w_{00}+w_{10}+w_{20} +w_{01}+w_{11}+ w_{21}).
\]
The nine Walsh coefficients described serve the same purpose as Walsh coefficients
in the biallelic case (see also the discussion section).

\subsection{Circuits for several alleles}

The sole circuit for a biallelic 2-locus system is
\[
w_{11}+w_{00}-w_{10}-w_{01}.
\]

There are in total 9 analogous circuits for a three-locus three-allele
system.
\[
w_{11}+w_{00}-w_{10}-w_{01}
\]

\[
w_{11}+w_{20}-w_{10}-w_{21}
\]

\[
w_{11}+w_{02}-w_{12}-w_{01}
\]

\[
w_{11}+w_{22}-w_{12}-w_{21}
\]

\[
w_{22}+w_{00}-w_{20}-w_{02}
\]

\[
w_{22}+w_{10}-w_{20}-w_{12}
\]

\[
w_{22}+w_{01}-w_{21}-w_{02}
\]

\[
w_{00}+w_{12}-w_{10}-w_{02}
\]

\[
w_{00}+w_{21}-w_{01}-w_{20}.
\]

\smallskip

However there are also six circuits of a different type:
\[
w_{00}+w_{11}+w_{22}-w_{01}-w_{12}-w_{20}
\]

\[
w_{00}+w_{11}+w_{22}-w_{02}-w_{10}-w_{21}
\]

\[
w_{00}+w_{12}+w_{21}-w_{01}-w_{10}-w_{22}
\]

\[
w_{00}+w_{12}+w_{21}-w_{02}-w_{11}-w_{20} 
\]

\[
w_{01}+w_{12}+w_{20}-w_{02}-w_{10}-w_{21} 
\]

\[
w_{01}+w_{10}+w_{22}-w_{02}-w_{11}-w_{20}.
\]

As mentioned, circuits are useful if
if one wants to analyze detailed aspects of
of gene interactions.
It has been established that
circuits have relevance for
recombination, or gene shuffling \citep{bps}.
For instance, suppose that
$
w_{00}+w_{11}>w_{01}+w_{10},
$
i.e., the circuit 
$
w_{00}+w_{11}-w_{01}-w_{10}>0
$
and that a population consists of
genotypes $10$ and $01$.
Then a single recombination event
\[
10+01 \mapsto 11+00
\]
can replace the poorly combined pair 
$\{ 10, 01 \}$ with a better
combined pair $\{00, 11 \}$.

For two loci and three alleles 
one needs to consider obstacles
that are not present in the biallelic
case.
In brief, if
\[
w_{00}+w_{11}+w_{22}>w_{01}+w_{12}+w_{20},
\]
then no single recombination event
for a $\{01, 12, 20\}$ population can
generate genotypes that are all
in the set $\{00, 11, 22\}$.
Circuits will not be discussed further here.

\newpage

\section{Two loci and four alleles}
In the remaining sections we will 
only discuss Walsh coefficients for interactions
since the lower order coefficients
are similar to the cases discussed.
For two loci and four alleles the
second order interactions 
are fairly similar to the case with 
two loci and three alleles.
Here one can compare $11$ with 
nine genotypes of Hamming distance two

\smallskip

$
w_{11}-w_{00}= (w_{10}-w_{00}) +(w_{01}-w_{00})
$

$
w_{11}-w_{20}= (w_{10}-w_{20}) +(w_{21}-w_{20})
$

$
w_{11}-w_{02}= (w_{12}-w_{02}) +(w_{01}-w_{02})
$

$
w_{11}-w_{22}= (w_{12}-w_{22}) +(w_{21}-w_{22})
$

$
w_{11}-w_{30}= (w_{10}-w_{30}) +(w_{31}-w_{30})
$

$
w_{11}-w_{03}= (w_{13}-w_{03}) +(w_{01}-w_{03})
$

$
w_{11}-w_{33}= (w_{13}-w_{33}) +(w_{31}-w_{33})
$

$
w_{11}-w_{23}= (w_{13}-w_{23}) +(w_{21}-w_{23})
$

$
w_{11}-w_{32}= (w_{12}-w_{32}) +(w_{31}-w_{32}).
$

\smallskip

From the expressons one concludes that additive fitness implies that

\[
w_{11}-\frac{1}{9} \left(w_{00} + w_{20} + w_{02}+  w_{22} + w_{30} + w_{03} + w_{33} + w_{23}+ w_{32} \right)=
\]

\[
\frac{1}{9} (3w_{10}+3w_{12}+3 w_{13}+3 w_{01}+3 w_{21}+3 w_{31}) +
\]

\[
\frac{1}{9} (2w_{00}+2w_{20}+2 w_{02}+ 2w_{22} +2w_{30}+2w_{03}+2 w_{23}+ 2 w_{32} + 2 w_{33}).
\]

\smallskip
It follows that the following form measures deviations from additive fitness
\[
w_{11}+\frac{1}{9} \left(w_{00} + w_{20} + w_{02}+  w_{22} + w_{30} + w_{03} + w_{33} + w_{23}+ w_{32} \right) \\
\]
\[
-\frac{1}{3} \, (w_{10}+w_{12}+w_{13}+w_{01}+w_{21}+w_{31}).
\]

By symmetry, for the entire set 
\[
S= \{ w_{11}, w_{12}, w_{13}, w_{21}, w_{22}, w_{23}, w_{31}, w_{32}, w_{33} \}
\]
one get similar forms.
The same is true for 
\[
T= \{ w_{00}, w_{01}, w_{02}, w_{03}, w_{10}, w_{20}, w_{30}  \}.
\]
However, the forms in $T$ are also linear combinations of the forms in $S$.
Following \citep{gs},
we define the nine forms obtained  from $S$ as the  second order Walsh coefficients.

\newpage

\section{Three loci and three alleles}
Recall that a second order interaction for the biallelic three-locus case
was obtained from the observation
\[
w_{111}-w_{001}=(w_{101}-w_{001})+(w_{011}-w_{001}).
\]
In this case the last coordinate is fixed at 1.
For the case with three alleles 
one can similarly compare $111$ with  $001$, but also with three  
other genotypes of  Hamming distance two such 
 that the third coordinate is fixed at 1 (i.e., $221, 201, 021$).
Additivity in the four cases implies

\[
w_{111}-w_{001}=w_{101}-w_{001} +  w_{011} - w_{001}
\]

\[
w_{111}-w_{221}=w_{121}-w_{221} +  w_{211} -w_{221}
\]

\[
w_{111}-w_{201}=w_{101}-w_{201}+ w_{211} -w_{201}
\]

\[
w_{111}-w_{021}=w_{121}-w_{021}+ w_{011} - w_{021}.
\]

\bigskip
By symmetry, a comparison of  $112$ 
and the genotypes $002$, $222$, $202$, $002$ 
gives

\[
w_{112}-w_{002}=w_{101}-w_{002} +  w_{011} -w_{002}
\]

\[
w_{112}-w_{222}=w_{122}-w_{222} +  w_{212} -w_{222}
\]

\[
w_{112}-w_{202}=w_{102}-w_{202}+ w_{212} -w_{202}
\]

\[
w_{112}-w_{022}=w_{122}-w_{022}+ w_{011} - w_{022},
\]

\bigskip
and  similarly from a comparison of  $112$ with $002$, $222$, $202$, $002$:

\[
w_{110}-w_{000}=w_{100}-w_{000} +  w_{010} -w_{000}
\]

\[
w_{110}-w_{220}=w_{120} - w_{220} +  w_{210} -w_{220}
\]

\[
w_{110}-w_{200}=w_{100}-w_{200}+ w_{210} -w_{200}
\]

\[
w_{110}-w_{020}=w_{120}-w_{020}+ w_{010} - w_{020}.
\]

From the 12 expressons one can 
determine an additive expectation for
\[
w_{110} + w_{111} +w_{112}.
\]
The expectation translates
to that the following form
measures deviations from
additivity:

\[
w_{110} + w_{111} +w_{112}
\]
\[
+\frac{1}{4} \left(w_{000}+w_{200}+w_{020}+ w_{220}+ w_{001} + w_{201} + w_{021}+  w_{221} + w_{002} + w_{202}+ w_{022} + w_{222}  \right)
\]
\[
-\frac{1}{2} ( w_{100} + w_{010} + w_{120}+w_{210} +w_{101} + w_{011} + w_{121}+w_{211}+ w_{102} + w_{012} + w_{122}+w_{212}).
\]

We denote the form $w_{11\ast}$
(as a short notation for a prediction based on
$
w_{110} + w_{111} +w_{112}
$).
With the same notation, one can describe
the second order Walsh coefficients as:
\[
w_{11 \ast}, \, \,  w_{12\ast}, \, \, w_{21\ast}, \, \, w_{22 \ast}, \, \,
w_{1 \ast 1}, \,  \, w_{1 \ast 2},  \,  \, w_{2 \ast 1},  \, \,   w_{2 \ast 2}, \, \,
w_{\ast 11},  \, \,   w_{\ast 12},  \, \,   w_{\ast 21},   \, \, w_{\ast 22}.
\]

\subsection{Third order interactions}
Similar to the biallelic case
a comparison of $111$ and $000$
gives 
\begin{equation}
w_{111}=w_{110}+w_{101}+w_{011}-w_{100}-w_{010}-w_{001}+w_{000}.
\end{equation}
However, for three alleles one can also compare $111$ with the other genotypes of Hamming distance
three, i.e.,  $222, 200, 020, 002, 220, 202, 022$ give

\begin{equation}
 w_{111}=w_{112}+w_{121}+w_{211}-w_{122}-w_{212}-w_{221}+w_{222}
\end{equation}

\begin{equation}
w_{111}=w_{110}+w_{101}+w_{211}-w_{100}-w_{210}-w_{201}+w_{200}
\end{equation}

\begin{equation}
w_{111}=w_{110}+w_{121}+w_{011}-w_{120}-w_{010}-w_{021}+w_{020}
\end{equation}

\begin{equation}
w_{111}=w_{112}+w_{101}+w_{011}-w_{102}-w_{012}-w_{001}+w_{002}
\end{equation}

\begin{equation}
w_{111}=w_{110}+w_{121}+w_{211}-w_{120}-w_{210}-w_{221}+w_{220}
\end{equation}

\begin{equation}
w_{111}=w_{112}+w_{121}+w_{211}-w_{102}-w_{212}-w_{201}+w_{202}
\end{equation}

\begin{equation}
w_{111}=w_{112}+w_{121}+w_{011}-w_{122}-w_{012}-w_{021}+w_{022}.
\end{equation}

From (17)-(24) a linear expectation on $w_{111}$ can be written as
\medskip
\[
w_{111}=\frac{1}{8}(w_{000}+w_{222}+w_{200}+w_{020}+w_{002}+w_{220}+w_{202}+w_{022}) 
\]
\[
-\frac{1}{4}(w_{100}+w_{122}+w_{120}+w_{102}+ w_{010}+w_{212}+w_{210}+w_{012}+w_{001}+w_{221}+w_{201}+w_{021}) 
\]

\[
+ \frac{1}{2}(w_{110}+w_{112}+w_{101}+w_{121}+ w_{011}+w_{211}) 
\]

Differently expressed, the following form measures third-order interactions

\[
w_{111}-\frac{1}{8}(w_{000}+w_{222}+w_{200}+w_{020}+w_{002}+w_{220}+w_{202}+w_{022}) 
\]
\[
+ \frac{1}{4}(w_{100}+w_{122}+w_{120}+w_{102}+ w_{010}+w_{212}+w_{210}+w_{012}+w_{001}+w_{221}+w_{201}+w_{021}) 
\]
\[
-\frac{1}{2}(w_{110}+w_{112}+w_{101}+w_{121}+ w_{011}+w_{211}).
\]

By symmetry, there are in total eight forms based on additive predictions
for 
\[
w_{111}, w_{112}, w_{121}, w_{112}, w_{221}, w_{212},w_{122}, w_{222}.
\]
Following \citet{gs} we define the eight forms as Walsh coefficients.
Obviously, one can obtain similar forms for $w_{ijk}$, where at least one member of 
the set $\{ i, j, k \}$ equals zero. However, such forms can also be described as linear combinations 
of the eight forms.

\section{discussion}
We have discussed interpretations of Walsh coefficients for biallelic systems.
Walsh coefficients of order two or higher can be interpreted as deviations
from additive expectations. 
The Walsh coefficients are built in a systematic way
from lower to higher order interactions. 
For instance, the Walsh coefficient of order three 
 (that measures three-way epistasis) for a biallelic three-locus system
is  based on the wild-type fitness $w_{000}$
the effects of changing single loci from 0 to 1, and 
three correction terms for second order
interactions. 

We have shown that Walsh-coefficients for an arbitrary number of loci
can be understood similarly, and provided explicit Walsh coefficients
for systems with two loci and three or four alleles, as well as for three loci
and three alleles. 
As demonstrated,  if one wants to define Walsh coefficients for  three or more alleles 
it is necessary to make some choices, whereas the biallelic case 
is canonical in principle. (Strictly speaking 
there is a choice also in the biallelic case, but
the impact will be restricted to the signs 
of the coefficients).

For an explicit  formula for 
Walsh coefficients for an arbitrary number
of alleles and related theory, see  \citet{gs}.
The properties of Walsh coefficients
in the general case are similar to the
biallelic case. The coefficients 
of order two or higher are zero if
fitness is additive, and all interactions
can be described as linear combinations
of the Walsh coefficients.
A recursive definition of Walsh coefficients
 for an arbitrary number of
 alleles can be found in \citep{mps}.

Walsh coefficients focus on average effects.
For a more detailed analysis of gene
interactions it can be useful to consider circuits.
For two loci and three alleles there are 15 circuits
(Section 2.1). It would be interesting with 
more results on circuits in the multiallelic setting.

 \end{document}